\documentclass[twocolumn,showpacs,aps,prb,amsmath,amssymb,floatfix]{revtex4}
\usepackage{graphicx}
\usepackage{epstopdf}
\usepackage{dcolumn}
\usepackage{bm}
\begin{document}
\title{Zero-energy Majorana states in a one-dimensional quantum wire with charge 
density wave instability}
\author{E. Nakhmedov$^{1,2}$ and O. Alekperov$^{2}$}
\affiliation{ $^1$Institut f\"ur Theoretische Physik,
Universit\"at W\"urzburg,
D-97074 W\"urzburg, Germany\\
$^2$Institute of Physics, Azerbaijan National Academy of
Sciences, H. Cavid ave. 33, AZ1143 Baku, Azerbaijan}
\date{\today}
\begin{abstract}
One-dimensional lattice with strong spin-orbit interactions (SOI) and Zeeman magnetic field is shown to lead 
to the formation of a helical charge-density wave (CDW) state near half-filling.
Interplay of the magnetic field,  SOI constants and the CDW gap seems to support Majorana bound states
under appropriate value of the external parameters. Explicit calculation of the quasi-particles' wave 
functions supports  a formation of the localized zero-energy state, bounded to the sample end-points.
Symmetry classification of the system is provided. 
Relative value of the density of states shows a precise zero-energy peak at the center of the band in 
the non-trivial topological regime. 
\end{abstract}
\maketitle
Recently, new exotic topological states of condensed matter, capable of 
supporting non-Abelian quasi-particle 
\cite{wilczek90}, have been suggested \cite{kitaev01, rg00, ivanov01, fk08}, which can be used as a 
fault-tolerant platform for topological quantum computation \cite{kitaev03, nssf08}. These topological phases
reveal chiral Majorana edge particles, being their own antiparticles, which are represented by the 
non-Abelian statistics with non-commutative fermionic exchange operators. 

Suggestion by Read and Green \cite{rg00}, that Majorana states can be realized at the vortex cores of a 
two-dimensional (2D) $p_x+ip_y$ superconductor, has  provoked new advances in engineering a semiconduction 
nanostructure with zero-energy state. Kitaev showed \cite{kitaev01} a possible realization of a single Majorana 
fermion at each end of a $p$-wave spinless superconducting wire. The effective $p$-wave superconductors were shown
\cite{stf09, slts10, alicea10, pl10} to be realized in a semiconductor film, in which $s$-wave pairing is induced by 
the proximity effect in the presence of spin-orbit interactions (SOI) and Zeeman magnetic field.

Formation of zero-energy Majorana bound states in one-dimensional ($1D$) quantum wire in the proximity 
to a s-wave superconductor and in the presence of SOI and the magnetic field has been argued recently 
in Refs. [\onlinecite{lss10,  oro10}]. 

In this Letter we predict a new realization mechanism of Majorana quasi-particles in a $1D$ crystal with charge 
density wave (CDW) instability.  We consider the  model of a $1D$ crystal around half-filling with 
strong spin-orbit interactions and in the presence of Zeeman magnetic field. There is an 
instability against formation of CDW and spin-density wave (SDW) in a such model.
The key to the quantum topological order is the coexistence of SOI with CDW or SDW state
and an externally induced Zeeman coupling of spins. We show that for the Zeeman coupling below a
critical value, the system is a nontopological CDW semiconductor. However, above the critical value of 
Zeeman field, the lowest energy excited state is a zero-energy Majorana fermion state for topological CDW crystal. 
Thus, the system is transmuted into a non-Abelian CDW state with increasing the external magnetic field. 

SDW and CDW are broken-symmetry ground states of highly anisotropic, so called quasi-$1D$ 
metals which are thought to arise as the consequence of electron-phonon or electron-electron 
interactions \cite{hkss88, gruner94}. These states have typical 1D character, and they can be conveniently 
discussed within the framework of various 1D models \cite{solyom79}.
CDW and SDW states are successfully realized in quasi-1D structures such as organic molecules of 
$(TMTSF)_2PF_6$, $(MDTTF)_2Au(CN)_2$, $(DMET)_2Au(CN)_2$, and $Au$, $In$, $Ge$ atomic wires grown by 
self-assembly on vicinal $Si(553)$, $Si(557)$ or $Ge(001)$ surfaces \cite{bsmm11, sw10}.

The model considered here is essentially 1D Hubbard model with on-site Coulomb interactions in the
presence of both Rashba and Dresselhaus SOI and Zeeman magnetic field. Non-interacting part $\hat{H}_0$ of the 
Hamiltonian $\hat{H}= \hat{H}_0 + \hat{H}_{int}$ in momentum space reads
\begin{eqnarray}
&&\hspace{-5mm}\hat{H}_0= \sum_{0<k<G/2} \sum_{\sigma, \sigma'}\bigg\{ \xi_k c_{k,\sigma}^{\dag} c_{k, \sigma'} 
\delta_{\sigma, \sigma'}+ \omega_Z  c_{k,\sigma}^{\dag}(\sigma_x)_{\sigma \sigma'} c_{k, \sigma'}+\nonumber\\
&&\hspace{-5mm}\alpha_R \sin (kd)c_{k,\sigma}^{\dag}(\sigma_z)_{\sigma \sigma'}c_{k, \sigma'} +
\alpha_D \sin (kd)c_{k,\sigma}^{\dag}(\sigma_y)_{\sigma \sigma'}c_{k, \sigma'}+\nonumber\\
&&\hspace{-3mm} (k \leftrightarrow k-G/2) \bigg\},
\label{bare-hamiltonian} 
\end{eqnarray}
where $\alpha_R$ and $\alpha_D$ are constants of Rashba and Dresselhaus SOI \cite{RD}, correspondingly, 
$\omega_Z=g \hbar \mu_B B/2$ is Zeeman energy of a
magnetic field $B$, $\xi_k= \epsilon_k-\mu$ with $\epsilon_k= -2t \cos (kd)$, and $\mu$ is the Fermi energy.
At half-filling $\mu=0$, and the electron-hole symmetry $\xi_{k-G/2}=-\xi_k$ for one-particle states is realized.  
$G=2 \pi/d$ is the reciprocal lattice vector with $d$ being the unit cell size. 
Interaction term $H_{int}$ in the Hamiltonian is written
\begin{eqnarray}
&&\hspace{-7mm}\hat{H}_{int}=\frac{1}{2 N}\sum_{0<q<G}\sum_{\sigma}\nonumber\\
&&\hspace{-7mm}\bigg \{ \sum_{-G/2<k<G/2-q\atop q-G/2<k'<G/2}U(k,k';q)
c_{k+q,\sigma}^{\dag}c_{k,\sigma}c_{k'-q, {-\sigma}}^{\dag}c_{k',{-\sigma}}+\nonumber\\  
&&\hspace{-7mm}\sum_{G/2-q<k<G/2 \atop -G/2<k'<q-G/2}U(k,k';q)
c_{k,\sigma}^{\dag}c_{k+q-G,\sigma}c_{k', {-\sigma}}^{\dag}c_{k'-q+G,{-\sigma}} \bigg \},
\end{eqnarray}
where $U$ is a strength of the Hubbard interaction and $N$ is the number of lattice sites. 
Note that the momentum summation in Eq. (\ref{bare-hamiltonian}) is taken over positive part of the Brillouin
zone, and the first four terms in Hamiltonian describe the right moving ($k>0$) particles. The left
moving ($k-G/2<0$) particles are taken into account by adding the 
terms with $k \leftrightarrow k-G/2$. 
The electron-hole order parameter at the density-wave instability is introduced as 
$\Delta_{\sigma}=\frac{V}{N}\sum_{0<k<G/2} \langle c_{k-G/2, \sigma}^{\dag}c_{k,\sigma} \rangle$
under assumption that $U(k,k',q)= V \delta (q-G/2)$. The complex-conjugate order parameter
is obtained by summing the electron-hole pairing over negative momentum part of the Brillouin zone,
$\Delta_{\sigma}^{\ast}=\frac{V}{N}\sum_{-G/2<k<0} \langle c_{k+G/2, \sigma}^{\dag}c_{k,\sigma} \rangle$.
CDW and SDW order parameters are defined as  $\Delta_{CDW} = \left( \Delta_{\uparrow}+ \Delta_{\downarrow}\right)/2$ and
$\Delta_{SDW} = \left( \Delta_{\uparrow}- \Delta_{\downarrow}\right)/2$, respectively.
Assuming $\Delta_{\uparrow} = \Delta_{\downarrow}$ for CDW, thereby we eliminate SDW ordering, and
$\Delta_{CDW} = \Delta_{\uparrow} = \Delta_{\downarrow}$.
For SDW we assume  $\Delta_{\uparrow} = -\Delta_{\downarrow}$, at the same time CDW formation is eliminated, and
$\Delta_{SDW} = \Delta_{\uparrow}= -\Delta_{\downarrow}$.
Further we use a common notation $\Delta$ for both CDW and SDW ordering, and replace
$\hat{H}_{int}$ in the mean field approximation by $\hat{H}_{int}^{MF}$
\begin{equation}
\hat{H}_{int}^{MF}=\sum_{0<k<G/2, \sigma}\bar{\sigma}\{\Delta c_{k,\sigma}^{\dag}c_{k-G/2,\sigma}+
\Delta^{\ast}c_{k-G/2, \sigma}^{\dag}c_{k,\sigma}\},
\label{int-MF}
\end{equation} 
where $\bar{\sigma}=1$ for CDW ordering, and $\bar{\sigma} =-\sigma= \mp 1$ for
SDW state. Hamiltonian $\hat{H}_{MF}=\hat{H}_0+\hat{H}_{int}^{MF}$ is written in the basis 
$\mathbf{\Psi}^{\dag}=\left( c_{k, \uparrow}^{\dag} c_{k, \downarrow}^{\dag}  c_{k-G/2, \downarrow}^{\dag}    
-c_{k-G/2, \uparrow}^{\dag} \right)$ as
\begin{equation}
\hat{H}_{MF}= \sum_{0<k<G/2} \{\mathbf{\Psi}^{\dag} \hat{\mathcal{H}} \mathbf{\Psi} +\xi_k+\xi_{-k+G/2} \}+
\frac{2}{V} |\Delta|^2,
\label{MF}
\end{equation}
with 
\begin{eqnarray}
&&\hat{\mathcal{H}}=\xi_k~ \tau_z \otimes \sigma_0 +\alpha_R \sin k~ \tau_0 \otimes \sigma_z +
\alpha_D \sin k ~\tau_z \otimes \sigma_y+\nonumber\\
&&{\omega}_Z ~\tau_z \otimes \sigma_x+\tau_j(\Delta)\otimes \sigma_j;
\label{Hamil-k}
\end{eqnarray}
where the Pauli matrices $\sigma$ and $\tau$ operate in spin and particle-hole spaces, $\otimes$ is the Kronecker
product of matrices. In the last term, $j=y$ for CDW and $j=x$ for SDW pairing 
\begin{eqnarray}
\tau_y(\Delta)=
\left( \begin{array}{cc}
0 &- i\Delta  \\
i\Delta^{\ast} & 0
\end{array} \right),
\quad {\rm and} \quad 
\tau_x(\Delta)=
\left( \begin{array}{cc}
0 & \Delta  \\
\Delta^{\ast} & 0
\end{array} \right),
\label{m-Delta}
\end{eqnarray}
The first term of Eq. (\ref{Hamil-k}) in the linearized form, $-\hbar \partial_y \tau_z$, with the 
third Zeeman term, $\omega_Z \sigma_x$ constitutes the massive Dirac equation. The charge density ordering, 
however, with the last term $\tau_j(\Delta) \sigma_j$ transforms the model to the four-band model.  

The pole of the single particle Green's function $G^{-1}(E,k)=E-\hat{\mathcal{H}}$ determines the quasiparticle 
energy
\begin{eqnarray}
&&E^2_{CDW}= \xi_k^2 +\alpha^2 \sin^2 k +|\Delta|^2+ \omega_Z^2 \pm \nonumber\\
&&\pm 2\sqrt{\xi_k^2 \alpha^2 \sin^2 k 
+ \omega_Z^2 |\Delta|^2 +\xi_k^2 \omega_Z^2};
\label{energy-spectrum}
\\
&&E_{SDW}^2=\left(|\xi_k| \pm \sqrt{\alpha^2 \sin^2 k + \omega_Z^2}\right)^2 + |\Delta|^2,
\label{energy-spectrum-SDW}
\end{eqnarray}
for CDW and SDW states, correspondingly. SO coupling constant $\alpha$ in the expressions for 
the energy spectrum is a renormalized
constant $\alpha = \sqrt{\alpha_R^2 + \alpha_D^2}$. Equation (\ref{energy-spectrum-SDW}) does not
allow a zero-energy mode due to a finite gap $\Delta$ at the origin. However, experimental evidences in many 
quasi-1D materials, e.g. in Bechgaard salt $(TMTSF)_2PF_6$ suggest a realization of an unconventional
SDW with an order parameter $\sim \Delta_1\sin k$ yielding a zero-energy state. The dispersive CDW or SDW gap 
can be derived from the extended Hubbard model with nonlocal interaction \cite{mdv08}. 
Further, we will discuss only the topological CDW state.

A small deviation from half-filling at $T=0$ was shown by Brazovskii et al. \cite{bgk80} to create 
a band of kink states within the Peierls gap. This picture is changed at finite temperatures.  
According to the phase diagrams in the temperature-chemical 
potential ($T, \mu$) and temperature-density ($T, n$) planes, callculated in Ref.[ \onlinecite{mf81}] on
the base of Brazovskii et al. theory \cite{bgk80}, for fixed electron density 
$1<n<n_L$, where $n_L$ is Leung's density \cite{leung75} at the triple point of the normal (N), 
commensurate (C) and incommensurate (IC) phases,  the kink band shrinks with increasing temperature 
until it vanishes at the IC-C transition. For  fixed temperature $0<T<T_L$  the kink band arises at 
some electron density $n>1$  and broadens with increasing density until the kinks become soft. At finite 
temperatures ($T<T_0$) and for small deviation of the chemical potential from half-filling 
$|\mu | < T_0 = 1.056 T_c(0)= (2/\pi)\Delta$, where 
$T_c(0)=(4We^{\gamma}/\pi)e^{-1/{\lambda}}$ is the transition temperature at $\mu=0$ \cite{mf81}, the system 
is in C-phase with vanishing mismatching between the electronic states $k$ and $G/2-k$ .
\begin{figure}
\resizebox{.48\textwidth}{!}{%
\includegraphics[width=1cm]{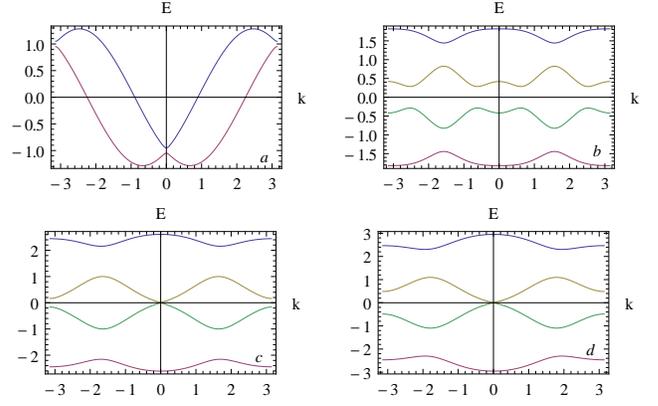}}
\caption {Energy spectrum is plotted according to Eq. (\ref{energy-spectrum}) for the fixed values of
$t=0.5$,  $\tilde{\alpha} =0.8$, and for the following
values of the dimensionless parameters, (a) $\tilde {\Delta}=0.0$,  
$\tilde {\omega}_Z=0.05$, $\tilde{\mu}=0.0$; (b) $\tilde {\Delta}=0.5$, $\tilde {\omega}_Z=0.7$,  
$\tilde{\mu}=0$; (c) $\tilde {\Delta}=0.7$,  
$\tilde {\omega}_Z=\sqrt{1.3}$, $\tilde{\mu}= - 0.1$; (d) $\tilde {\Delta}=0.7$, $\tilde {\omega}_Z=\sqrt{2.18}$,
$\tilde{\mu}=-0.3$}.  
\label{Band-R}
\end{figure}

Solution of the energy spectrum for different values of  $\tilde {\alpha}$, $\tilde{\mu}$, 
$\tilde {\Delta}$, and $\tilde {\omega}_Z$ is plotted in Fig. \ref{Band-R}, where the dimensionless 
parameters with tilde are given in the unit of the halved band width $2t$.
Solution of Eq. (\ref{energy-spectrum}) $\alpha_R=\alpha_D=\Delta=\omega_Z$ yields a usual 
cosine-band in the reduced Brillouin zone. 
SOI results in two shifted cosine-bands along $k$-axes, whereas Zeeman splitting doubles 
the band along the energy axes, opening a gap at the anticrossing point (see, Fig. \ref{Band-R}a). 
Formation of the density wave opens a 
gap at the boundary of the Brillouin zone. 

The energy spectrum at the center of the Brillouin zone for the topological CDW with gapped ``bulk'' 
states and zero-energy end-states can be written as
\begin{equation}
E_{CDW}^{(0)}=E(0)=\Big|\omega_Z - \sqrt {\mu_t^2 + |\Delta|^2}~\Big|,
\label{zero}
\end{equation}  
where $\mu_t= -2t-\mu$.
A magnetic-field dominated gap at the center of the band for  $\omega_Z^2>|\Delta|^2+\mu_t^2$ turns to 
the pairing-dominated one for  $\omega_Z^2<|\Delta|^2+\mu_t^2$, (Figs. \ref{Band-R}d and b, correspondingly). 
A quantum phase transition from topological non-trivial to trivial phase  occurs at 
$\omega_Z^2=|\Delta|^2+\mu_t^2$. The gap at $k=0$ 
vanishes under this condition emerging Majorana fermion states at the ends of the wire, which is plotted
in Fig. \ref{Band-R} for the dimensionless parameters $\tilde {\alpha}=0.8$, $\tilde {\Delta}=0.7$, 
$\tilde {\omega}_Z=\sqrt{1.3}$, and $\tilde{\mu}= - 0.1$. 

It is possible to check that the Hamiltonian $\hat{\mathcal{H}}$ respects 
time-reversal symmetry (TRS) $U_T\hat{\mathcal{H}}^{\ast}(k) U_T^{-1}= \hat{\mathcal{H}}(-k)$ with TRS 
operator $T=U_TK$ 
in the absence of the magnetic field, and particle-hole symmetry (PHS) 
$U_P \hat{\mathcal{H}}^{\ast}(k) U_P^{-1} = -\hat{\mathcal{H}}(-k)$ with PHS operator $P=U_P K$. 
Here, $K$ is the complex conjugate operator, 
$U_T=\sigma_0 \otimes i \sigma_y$ and $U_P=\sigma_x \otimes \sigma_0$ satisfying $T^2=-1$ and $P^2=1$. 
The TRS operator transforms $k \to -k$ as well as $c_{k \uparrow} \Leftrightarrow c_{k \downarrow}^{\dag}$
and $c_{k \downarrow} \Leftrightarrow  - c_{k \uparrow}^{\dag}$, resulting in 
$\Delta \leftrightarrow \Delta^{\ast}$ for the order parameter  and keeping the excitation spectrum 
unchanged $\xi_{-k}=\xi_k$. Instead, the PHS operator transforms 
\begin{equation}
c_{k \uparrow} \Leftrightarrow c_{k-G/2 \downarrow}^{\dag} \qquad {\rm and} \qquad 
c_{k \downarrow} \Leftrightarrow  - c_{k-G/2 \uparrow}^{\dag},
\label{PHS}
\end{equation}
keeping unchanged the order parameter $\Delta$.   
PHS entails an energy spectrum symmetric about the Fermi level.
According to symmetry classification the system belongs to $DIII$ class which can be topologically 
nontrivial \cite{srfl08} provided that both TRS and PHS are satisfied. 
An external magnetic field breaks TRS and drives the system from $DIII$ to 
$D$ class, which possesses a single Majorana zero-energy mode at each end of the wire.

The main feature of Majorana fermion is that it is own 'anti-particle'. This property can be proved
for a $1D$ unconventional CDW model \cite{mdv08, hkss88} with dispersive and complex order
parameter $\Delta_k=\Delta_0 \sin(kd)$ by mapping it to the Kitaev's model \cite{kitaev01} for 
the p-wave superconductor. Hamiltonian of a $1D$ unconventional CDW model becomes invariant under 
the particle-hole transformations 
$c_k^v \equiv c_k \leftrightarrow c_{k}^{c \dag} \equiv c_{k-G/2}^{\dag}$ and 
$c_k^{v \dag} \leftrightarrow c_k^c$ in momentum space or $d_n^v \leftrightarrow  d_n^{c \dag}$ and   
$d_n^{v \dag} \leftrightarrow d_n^{c}$ in site-representation, where the spin index is neglected due to
the spin degeneration. The PHS transforms it to the Kitaev's one 
\begin{eqnarray}
&&\hat{H}_0=\sum_n \{ -2t(d_n^{v \dag}d_{n+1}^v+d_{n+1}^{v \dag}d_n^v) +2i\Delta_0d_n^{v \dag}d_{n+1}^{v \dag}+\nonumber\\
&&2i\Delta_0^{\ast} d_{n+1}^{v} d_{n}^{v}\},
\label{kitaev}
\end{eqnarray}
which reveals  the Majorana end states. It is easy to show that the PHS conditions (\ref{PHS}) transform 
our Hamiltonian (\ref{bare-hamiltonian}) and (\ref{int-MF}) to the form, describing the $s$-wave type 
superconductor  with misaligned spins but with the same momenta $k$ of Cooper pairs, which should reveal 
again the Majorana quasi-particles.

 Majorana bound states arise at the interface of trivial and
topological regions under certain condition by varying the parameters of $1D$ wire. In order to understand 
the localized character of the zero energy state we rewrite Hamiltonian in the real coordinate space. 
We linearize the cosine energy spectrum around the Fermi level $k_F=G/4$ as 
$\xi_k=\epsilon_k-\mu=4t \sin \frac{(k+k_F)d}{2}\sin \frac{(k-k_F)d}{2} \approx v_F \hbar(k-k_F)
 \to v_F \hbar( -i\frac{\partial}{\partial y}-k_F)$ for right-mover,
and $\xi_{k-G/2}\approx -v_F \hbar (k+k_F) \to -v_F \hbar (i\frac{\partial}{\partial y}-k_F)$ 
for left-mover, and the SO coupling term 
$\sin (dk) \to -id \frac{\partial}{\partial y}$. One can see that $\mu_z=v_Fk_F\hbar$; at half-filling
$\mu=0$ and $\mu_t=v_Fk_F\hbar=2t$. 
Schr\"odinger equation, corresponding to zero energy, reads
\begin{eqnarray}
&&\hspace{-5mm}\left[-\mu_t - i(v_F \hbar + \nu_{\sigma} \alpha_R) \frac{\partial}{\partial y} \right]\psi_{\sigma}^R 
+\left(\omega_Z -\nu_{\sigma} \alpha_D
\frac{\partial}{\partial y}\right)\psi_{-\sigma}^R+\nonumber\\
&&\hspace{-3mm}\Delta \psi_{\sigma}^L=0\nonumber\\
&&\hspace{-5mm}\left[\mu_t + i(v_F\hbar + \nu_{\sigma} \alpha_R) \frac{\partial}{\partial y}\right]\psi_{\sigma}^L 
+\left(\omega_Z + \nu_{\sigma} \alpha_D
\frac{\partial}{\partial y}\right)\psi_{-\sigma}^L +\nonumber\\
&&\hspace{-3mm}\Delta^{\ast}\psi_{\sigma}^R=0,
\end{eqnarray}
where $-\sigma=\downarrow, \uparrow$, and  $\nu_{\sigma}=\pm$ for  $\sigma=\uparrow, \downarrow$, correspondingly.
For long enough wire $L \gg 1$, we choose the magnetic field $\omega_Z^2<\mu_t^2+|\Delta|^2$ for $y \in [0,L]$
and $\omega_Z^2>\mu_t^2+|\Delta|^2$ outside this interval.
By choosing the wave functions $\Psi^T(y)=\exp \{iky\} (b_{\uparrow}^R,~b_{\downarrow}^R,~b_{\downarrow}^L,~
-b_{\uparrow}^L)^T$, 
one gets the determinant 
equation $det |\bar{\mathbf{H}}|=0$ to find $k$, where
\begin{eqnarray}
&&\mathbf{H}=v_F\hbar(k-k_F)~\tau_z \otimes \sigma_0+\alpha_R k~\tau_0 \otimes \sigma_z+\nonumber\\
&&\omega_Z~\tau_z \otimes \sigma_x+ \alpha_Dk~\tau_z \otimes \sigma_y +\Delta~\sigma_y \otimes \tau_y.
\label{hamil-linear}
\end{eqnarray}
The allowed values of $k$ are obtained from the equation, 
$(v_F^2 \hbar^2 -\alpha^2)k^2-2k(\mu_t v_F \hbar \pm i|\Delta|\alpha)-\mathcal{L}=0$, 
where $\mathcal{L}=\omega_Z^2-|\Delta|^2-\mu_t^2$. For $\mathcal{L} = 0$ this equation has a real root $k=0$, 
corresponding to a single allowed state in the gap. Since there is no other state for a quasiparticle 
to move, this 
state is localized and it seems to be protected against local perturbations. 
For $\mathcal{L} \ne 0$, $k$ takes complex values, signaling on realization of a gapped state. 
In this case the wave function decays exponentially in both sides of $y=0$ but with 
different localization lengths. General solution for $k$ reads
\begin{equation}
k_{\nu}=\frac{k_F \pm i |\bar{\Delta}| \bar{\alpha} +\nu \sqrt{(k_F\bar{\alpha} \pm i |\bar{\Delta}|)^2+
{\bar\omega}_Z^2(1-{\bar{\alpha}}^2)}}{1 -{\bar{\alpha}}^2},
\end{equation}
where $\bar{\alpha}=\frac{\alpha}{v_F\hbar}$, $\bar{\Delta}=\frac{\Delta}{v_F\hbar}$, 
$\bar{\omega}_Z=\frac{\omega_Z}{v_F \hbar}$, and $\nu = \pm$. The wave function decays exponentially if,
 generally speaking $|\Delta|$, $\alpha \ne 0$. For 
$\alpha = 0$, $k_{\pm}=\frac{\mu_t \pm \sqrt{\omega_Z^2 - |\Delta|^2}}{v_F \hbar}$ and the trivial CDW 
state is gapped for $\omega_Z < |\Delta|$, which is destroyed for $\omega_Z > |\Delta|$. 

Majorana bound state is formed by varying the parameters $\Delta$, $\omega_Z$, and $\mu$.
We consider a linearized Hamiltonian, Eq. (\ref{hamil-linear}), for the relevant momenta near $k=0$ and $\mu_t=0$,
assuming a spatial variation of the magnetic field $\omega_Z=\Delta + by$ near $y=0$, which crosses a constant 
gap $\Delta >0$. For simplicity, Dresselhaus SOI is neglected, $\alpha_D=0$, and $\Delta$ is chosen to be real.  
Following Oreg et al. \cite{oro10}, the squared, due to the particle-hole symmetry, Hamiltonian
(\ref{hamil-linear}), $\mathbf{H}^2$, is reduced to the diagonal form by mean of the unitary operator
$U=\frac{1}{2}(\tau_z + i\tau_y+i\sigma_x \tau_z + \sigma_x \tau_y)$,
\begin{equation}
\tilde{\mathbf{H}}=U{\mathbf{H}^2}U^{\dag}=[\omega_Z^2 +\Delta^2 + (\alpha_R k)^2]-\alpha_R \hbar b \sigma_z \tau_z
+2\omega_Z\sigma_z \tau_0,
\end{equation}
with spectrum $E^2=(\omega_Z \pm \Delta)^2 \pm \alpha_R \hbar b$.
The term, proportional to $b \sigma_z$, appears in Hamiltonian due to the topological defect at the ends
of the wire, which bridges two edges of the conduction and valence bands. The bound state may form if $\Delta$
varies in space and crosses $\omega_z$.
\begin{figure}
\resizebox{.48\textwidth}{!}{%
\includegraphics[width=1cm]{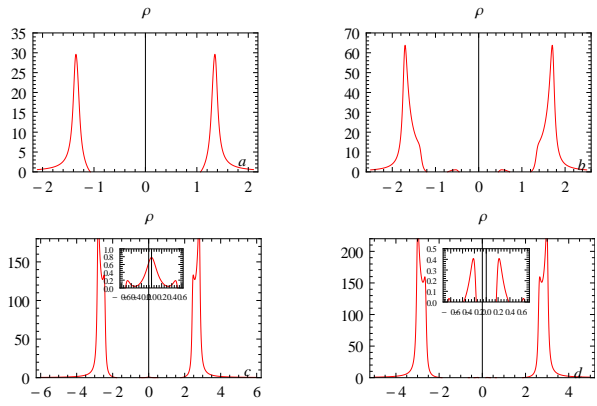}}
\caption {The relative change in the DOS $\delta \rho(\epsilon, V)/\rho^{(0)}(\epsilon)$ for 
(a) $\tilde{\alpha}=0.8$, $\tilde{\Delta} = 0.0$ and $\tilde{\omega}_Z=0.3$, 
(b) $\tilde{\alpha}_R=0.6$, $\tilde{\Delta} = 0.7$ and $\tilde{\omega}_Z=0.5$, (c) $\tilde{\alpha}_R=0.3$, 
$\tilde{\Delta} = 1.0$ and $\tilde{\omega}_Z=\sqrt{2.0}$, and (d)  $\tilde{\alpha}_R=0.3$, $\tilde{\Delta} = 1.0$ 
and $\tilde{\omega}_Z=1.6$. The inelastic scattering rate is chosen 
to be $\tilde{\eta}=0.05$. Inset in (c) shows a zero-energy peak, corresponding to Majorana quasi particle, which 
disappears in inset (d) by destroying the condition.}  
\label{DeltaDOS-R}
\end{figure} 

Zero-energy Majorana state in the Peierls gap can be experimentally detected from the tunneling experiments, 
where the conductivity of  the tunneling contact is expressed through the one-particle density 
of states (DOS), $\rho (\epsilon, T)$, as
\begin{equation}
\frac{\delta G(V,T)}{G^{(0)}}=\int_{-\infty}^{+\infty} \frac{d\epsilon}{4T} ~\frac{\delta \rho (\epsilon)}{\rho^{(0)}}
\bigg[\frac{1}{\cosh^2 \frac{\epsilon - eV}{2T}} + \frac{1}{\cosh^2 \frac{\epsilon + eV}{2T}}\bigg].
\label{tun.cur}
\end{equation}
At $T=0$ this expression is written
$\delta G(\epsilon)/G^{(0)}=[\rho(\epsilon,0)-\rho^{(0)}]/\rho^{(0)}= \delta \rho(\epsilon)/\rho^{(0)}$,
where $\rho^{(0)}$ is the DOS of a pure system. 
The DOS is found from the conventional expression
$\rho(\epsilon)=\int_{-\pi}^{\pi} \frac{dk}{2\pi}\sum_n \delta(\epsilon -E_n(k))$,
where $E_n(k)$ is the energy spectrum for $n=1,2,3,4$ given by Eq. (\ref{energy-spectrum}).
The delta-function is regularized for numerical calculations, replacing it by Lorenzian function
$\delta (\epsilon - E_n(k))=\eta/\{(\epsilon-E_n(k))^2+\eta^2\}$, where $\eta$ is the rate of inelastic processes. 
A formation of the Majorana quasi-particle in the center of the band is clearly seen in the relative value 
of the DOS $\delta \rho(\epsilon)/\rho^{(0)}$. Evolution of the central peak in 
$\delta \rho (\epsilon)/\rho^{(0)}$ is depicted in Fig. \ref{DeltaDOS-R}, where the central 
peak emerges only for special values of the external parameters satisfying the critical condition 
$\omega_Z^2=|\Delta|^2+\mu_t^2$.  
Note that midgap states have been observed recently in a topological superconducting phase by Mourik et al. 
\cite{mzfp12} and by Das et al. \cite{drmo12} in zero-bias measurements on $InSb$ and $InAs$ nanowires, 
contacted with one normal (gold) and one superconducting electrode.

An artificial string of $Au$, $In$, $Ge$, $Pb$ atoms on vicinal $Si(557)$, $Si(553)$ and 
$Ge(001)$ surfaces seems to be suitable for experimental realizations. These structures with a large 
lateral chain spacing ($\sim 1.6 nm$)  can be built \cite{nwh02} by placing 
metallic atoms side-by-side on a non-conducting template by using e.g. a scanning tunneling microscope. 
Angle-resolved photoemission data indicates a $1D$ 
electron pocket with very weak transverse dispersion in these structures. The ratio of the parallel
and transverse hopping integrals $t_{\|}/t_{\perp}$ was determined from a tight-binding fit to the Fermi contour 
to be larger than $60$ \cite{sw10}. Therefore, the structures are three-dimensional with practically in-wire
motion of particles.
These structures exhibit a Peierls instability below $\sim 150-200 K$. Recently, a spin polarized CDW has 
been observed \cite{tlps12} in $Pb/Si(557)$, where the Fermi surface nested charge density instability 
occurs  by appropriate choice of band- filling, spin-orbit coupling and external parameters.  
The Rashba parameter in this structure was found to be  1.9 eV $\AA$ for the value of the Rashba 
splitting $0.2 \AA^{-1}$. High values of the band gap and the SOI constants may allow  to realize a 
topological CDW phase at higher temperatures, making a significant step compared to previous
mechanism  to detect Majorana state in topological superconductors.

We showed in this paper a possible realization of zero- energy Majorana state in 
the CDW phase of a 1D crystal. CDW state in 1D crystal is realized due to nesting of the Fermi level. 
The wave function of this state  ``mixes'' an electron state $\psi_{k, \sigma}$ with a momentum $k>0$ 
above the Fermi level with an hole state $\psi_{k-G/2, \sigma}$ with a momentum $k-G/2<0$ below the Fermi 
level, which resembles the Bogolyubov-de Gennes wave 
function with mixed electron and hole states too. A quasiparticle excitation in the topological CDW state 
emerges as a localized zero- energy state in the middle of the Brillouin zone. Since CDW phase is realized at 
higher temperatures, this new mechanism facilitates an observation of Majorana particles and their 
implementation for  the quantum computations.   

E.P. would like to acknowledge B. Trauzettel and J. C. Budich for valuable discussions.
This work was supported by the Scientific Development Foundation of the Azerbaijan Republic under
Grant Nr. EIF-2012-2(6)-39/01/1.


\begin{thebibliography}{99}
\bibitem{RD} Rashba and Dresselhaus SOI, resulting from the bulk inversion  asymmetry and structural 
inversion asymmetry, correspondingly, are expressed in a $2D$  $\{yz\}$ plane as 
$H_R=\alpha_R ( p_y \sigma_z -p_z \sigma_y)$ and $H_D=\alpha_D(p_y \sigma_y - p_z\sigma_z)$, which are 
reduced to the forms  of $H_R=\alpha_R p_y \sigma_z$ and $H_D= p_y \sigma_y$ for a $1D$ wire lying along 
$y$-direction.
\end{thebibliography}
\end{document}